\documentclass[preprint2]{aastex}

\shorttitle{Traveling wave MRI}
\shortauthors{R\"udiger et al.}

\begin{document}

\title{The traveling wave MRI in cylindrical Taylor-Couette flow:
        comparing wavelengths and speeds in theory and experiment}

\author{G\"unther R\"udiger}
\affil{Astrophysikalisches Institut Potsdam, An der Sternwarte 16,
D-14482 Potsdam, Germany\\gruediger@aip.de}

\author{Rainer Hollerbach}
\affil{Department of Applied Mathematics, University of Leeds,
Leeds, LS2 9JT, United Kingdom\\rh@maths.leeds.ac.uk}

\author{Frank Stefani, Thomas Gundrum, Gunter Gerbeth}
\affil{Forschungszentrum Rossendorf, P.O. Box 510119, D-01314 Dresden,
Germany\\F.Stefani@fz-rossendorf.de, Th.Gundrum@fz-rossendorf.de,
G.Gerbeth@fz-rossendorf.de}

\author{Robert Rosner}
\affil{Department of Astronomy and Astrophysics, University of Chicago,
Chicago, IL 60637, USA\\rrosner@astro.uchicago.edu}

\begin{abstract}
We study experimentally the flow of a liquid metal confined between
differentially rotating cylinders, in the presence of externally
imposed axial and azimuthal magnetic fields.  For increasingly large
azimuthal fields a wave-like disturbance arises, traveling along
the axis of the cylinders.  The wavelengths and speeds of these
structures, as well as the field strengths and rotation rates at
which they arise, are broadly consistent with theoretical predictions
of such a traveling wave magnetorotational instability.
\end{abstract}


\section{Introduction}

The magnetorotational instability (MRI) arises in a broad range of
astrophysical problems, most importantly in accretion disks, where
it is generally accepted to be the source of the turbulence needed
for material to spiral inward and accrete onto the central
object \citep{BH91}.  Because of this crucial role that it plays in
astrophysics, there is considerable interest in trying to study the
MRI in the laboratory \citep{RRB04}.  One recent suggestion
\citep{HR05,RHS05} involves externally imposing combined axial and
azimuthal magnetic fields, which yields a new type of traveling
wave MRI.  In this letter we present experimental evidence of these
traveling waves, in accordance with the theory.

The MRI is a mechanism whereby a differential rotation flow that
satisfies the Rayleigh criterion, and is therefore hydrodynamically
stable, may nevertheless be magnetohydrodynamically unstable.  The
addition of a magnetic field allows angular momentum to be transferred
outward by the magnetic tension in the field lines, thereby bypassing
the Rayleigh criterion, which relies on individual fluid parcels
conserving their angular momentum.  The MRI is particularly relevant
to Keplerian flows such as those found in accretion disks, where
$\Omega\sim r^{-3/2}$, which would indeed be stable according to the
Rayleigh criterion.

The simplest design for attempting to reproduce the MRI
in the lab is based on the familiar Taylor-Couette problem,
consisting of the flow between differentially rotating cylinders.
While one cannot achieve precisely a Keplerian flow profile in this
problem, by appropriately choosing the rotation rates of the inner and
outer cylinders one can easily achieve a Rayleigh-stable profile,
which is all that is really required.  By suitably adjusting the
strength of an externally imposed magnetic field one should then be
able to destabilize the flow again, via the MRI.

Unfortunately, the situation is not quite so simple after all.  If
the imposed field is purely axial \citep{RZ01,JGK01}, the relevant
parameter for the onset of the MRI turns out to be the magnetic Reynolds
number $Rm=\Omega_{\rm i} r_{\rm i}^2/\eta$, which must exceed $O(10)$.
The hydrodynamic Reynolds number $Re=\Omega_{\rm i} r_{\rm i}^2/\nu$ then
exceeds $O(10^6)$, due to the extremely small magnetic Prandtl numbers
$Pm=\nu/\eta$ of liquid metals ($\nu$ is the viscosity, $\eta$ the
magnetic diffusivity).  Such large rotation rates can be achieved,
but for increasingly large Reynolds numbers end-effects become
increasingly important, and at $Re\gtrsim O(10^6)$ may well disrupt
the experiment \citep{HF04}.

In contrast, if a combined axial and azimuthal field is imposed
\citep{HR05,RHS05}, the relevant parameter turns out to be $Re$,
which must only be as large as $O(10^3)$ to obtain the MRI.  These
end-effects are therefore less severe.  The solutions in this case are
also somewhat different from those for purely axial imposed fields; one
obtains much the same Taylor vortices as before, but the whole pattern
now drifts along the length of the cylinders \citep{Knobloch}.  This
unfortunately introduces significant end-effects of its own, familiar in
other contexts, such as drifting dynamo waves \citep{TPK98}.  Nevertheless,
we will see that the experimental results in a bounded cylinder agree
reasonably well with the theoretical results in an unbounded cylinder.

\section{Experimental Results}

The experimental apparatus consists of a cylindrical annulus made of
copper, with $r_{\rm i}=4$ cm, $r_{\rm o}=8$ cm, and height 40 cm.  The top
endplate is made of plexiglass, and is stationary; the bottom endplate is
made of copper, and rotates with the outer cylinder.  An axial magnetic
field is imposed by running a current, up to 200 A, through a series of
coils surrounding the entire apparatus; an azimuthal magnetic field is
imposed by running a current, up to 8000 A, through a rod along the central
axis.  Field strengths of several hundred G can then be achieved, for both
$B_z$ and $B_\phi$.  The fluid contained within the vessel is a GaInSn alloy,
having density $\rho=6.4\ \rm g/cm^3$, viscosity $\nu=3.4\cdot 10^{-3}\ \rm
cm^2/s$ and magnetic diffusivity $\eta=2.4\cdot10^3\ \rm cm^2/s$, so $Pm=1.4
\cdot10^{-6}$.  Measurements were made by two ultrasonic transducers mounted
on opposite sides of the top endplate, 1.5 cm from the outer wall.  These
provided measurements of $v_z$ at these particular locations, along the
entire 40 cm depth of the container.  See also \citet{SGGRSSH} for further
details of the experimental setup.

For the results presented here, the rotation rates of the inner and
outer cylinders were fixed at $\Omega_{\rm i}=0.377\ \rm s^{-1}$ and
$\Omega_{\rm o}=0.102\ \rm s^{-1}$, so $\Omega_{\rm o}/\Omega_{\rm i}
=0.27$, within the Rayleigh-stable regime $\Omega_{\rm o}/\Omega_{\rm i}>
(r_{\rm i}/r_{\rm o})^2=0.25$.  The Reynolds number $Re=\Omega_{\rm i}
r_{\rm i}^2/\nu=1775$.  The axial field $B_z$ was also fixed, at 77.2 G,
corresponding to a Hartmann number $Ha=B_z r_{\rm i}/\sqrt{\rho\mu\eta\nu}
=12$.  The azimuthal field $B_\phi$ was then varied, between 0 and 350 G
(at $r_{\rm i}$), corresponding to electric currents up to 7000 A along the
central rod.

%%%%%%%%%%%%%%%%%%%%%%%%%%%%%%%%%%%%%%%%%%%%%%%%%

\begin{figure}
\epsscale{0.95}
\plotone{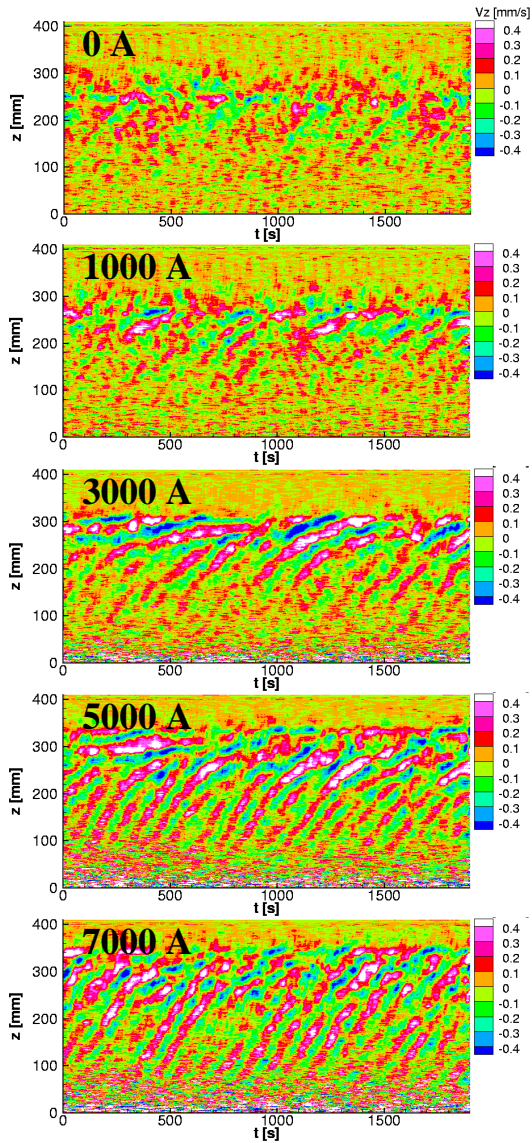}
\caption{Ultrasound measurements of the axial velocity $v_z$, with the
time-average removed.  Time is plotted on the horizontal axis; height
along the cylinder on the vertical axis.  The current along the central
rod varies from 0 to 7 kA as indicated, corresponding to
$B_\phi(r_{\rm i})$ up to 350 G.}
\end{figure}

\begin{figure}
\epsscale{0.95}
\plotone{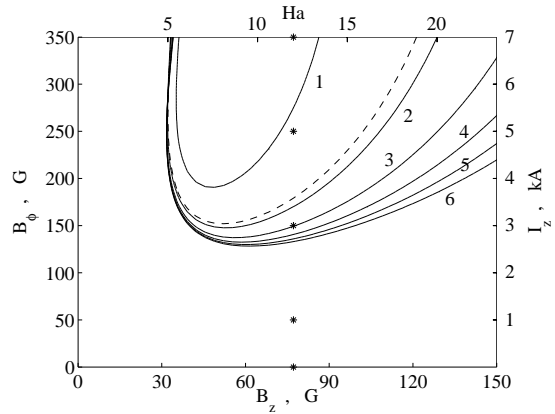}
\caption{Contour plot of $Re_c$ as a function of $B_z$ and $B_\phi$.  On
the horizontal axes $B_z$ is given both in G, and in nondimensional form,
as $Ha=0.1555\cdot B_z/\rm G$.  On the vertical axes $B_\phi(r_{\rm i})$
is given both in G, and in terms of the required axial current $I_z$.  The
numbers beside individual contour lines denote values from 1000 to 6000;
the dashed contour line is 1775, the value used in the experiment.  The
five asterisks correspond to the five experimental runs in Fig.\ 1.}
\end{figure}

\begin{figure}
\epsscale{0.95}
\plotone{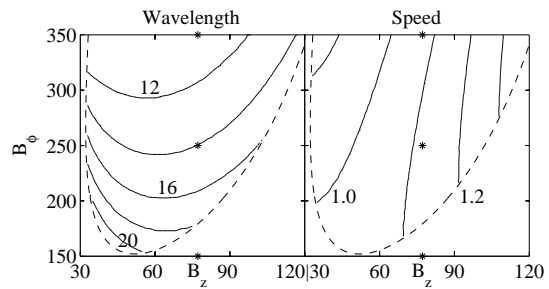}
\caption{The left panel shows the wavelengths, in cm, the right panel
shows the speeds, in mm/s, of the solutions at $Re=1775$.  As in Fig.\ 2, the
dashed curves are the stability boundary at this particular $Re$.  The three
asterisks correspond to the three experimental runs at 3, 5 and 7 kA,
where once again the wavelengths were approximately 8, 6 and 6 cm,
respectively, and the speeds 0.6, 0.8 and 0.8 mm/s.}
\end{figure}

%%%%%%%%%%%%%%%%%%%%%%%%%%%%%%%%%%%%%%%%%%%%%%%%%

Figure 1 shows the temporal variation of the axial velocity $v_z$, as a
function of $z$ along the container.  The (depth-dependent) time-average
has been subtracted, to remove the two Ekman vortices induced by the
endplates \citep{KJG04,HF04}.  As $B_\phi$ is increased, we see the gradual
emergence of ever more coherent structures, drifting at speeds of 0.5--1
mm/s.  Focussing attention on $z$ between 10 and 20 cm, where the waves are
most clearly defined, we obtain around 0.6 mm/s for 3 kA, and 0.8 mm/s for
5 and 7 kA.  The corresponding wavelengths are 7--8 cm for 3 kA, and 5-6 cm
for 5 and 7 kA.  We claim that these structures are precisely the expected
traveling wave MRI.

Note though that end-effects do indeed play an important role in Fig.\ 1,
for example at the upper boundary, where the waves die away some 5--10 cm
from the end.  Another indication of the importance of end-effects,
and in particular the asymmetry between the two ends, can be seen in the
0 A results: the asymmetry between top and bottom that is already visible
even in this case must be caused entirely by the endplates, as a purely
axial field does not distinguish between $\pm z$.

\section{Theoretical Analysis}

Figure 2 shows the critical Reynolds number for the onset of this traveling
wave MRI in an unbounded cylinder, as a function of the externally imposed
fields $B_z$ and $B_\phi$; these results were computed as in \citet{HR05} or
\citet{RHS05}, with conducting boundary conditions.  Provided $B_z>30$ G and
$B_\phi\gtrsim150$ G, Reynolds numbers of $O(10^3)$ are already sufficient.
If $B_\phi$ is less than 150 G, $Re_c$ gradually rises, until for $B_\phi=0$
we would have $Re_c>O(10^6)$, the familiar result from the analysis of purely
axial fields.  In contrast, if $B_z$ is less than 30 G, the instability
simply ceases to exist; that is, $Re_c$ has an essentially vertical asymptote
at this boundary.

The five asterisks in Fig.\ 2 correspond to the five plots shown in Fig.\ 1,
and the dotted contour line to the experimental value $Re=1775$.  We see
therefore that the experimental runs with $I_z=0$, 1 and 3 kA should be
stable, whereas the ones with 5 and 7 kA should be unstable.  This is
broadly in agreement with Fig.\ 1, although there even the supposedly stable
runs already show hints of traveling wave disturbances, particularly at 3
kA.  However, the waves are much more strongly developed for the 5 and 7 kA
runs, as predicted by Fig.\ 2.

Fig.\ 3 shows the wavelengths and speeds of the unstable solutions at $Re=
1775$.  Wavelengths are in the range 10--20 cm, speeds around 1.1 mm/s (for
$B_z=77.2$ G).  The speeds agree rather well with Fig.\ 1; the experimental
result that 5 and 7 kA yield virtually the same speed, 0.8 mm/s, also
nicely matches the theoretical prediction that the speed should be almost
independent of $B_\phi$ (provided $B_\phi$ is large enough to be in the
unstable regime at all).  The wavelengths do not agree quite so well; the
values in Fig.\ 1 are barely half those in Fig.\ 3.  Presumably this is
again due to end-effects; a 20 cm long wave certainly could not traverse a
40 cm long container without experiencing significant end-effects.

\section{Conclusion}

In this letter we have presented experimental evidence for the existence of
traveling wave disturbances in a liquid metal Taylor-Couette apparatus, and
shown them to be in reasonable agreement with the theoretical predictions,
particularly with regard to the wave speeds.  Future experimental work will
more thoroughly map out the entire $B_z$, $B_\phi$, $\Omega_{\rm i}$ and
$\Omega_{\rm o}$ parameter space, as well as explore the role that different
axial boundary conditions, on both the flow \citep{KJG04} and the field
\citep{LIU06}, might play.  Future numerical work will similarly consider
the problem in bounded cylinders, with different boundary conditions on the
endplates.  Only then will we fully understand what influence the endplates,
and possible asymmetries between them, have on the MRI.

\acknowledgments

This work was supported by the German Leibniz Gemeinschaft, under program
SAW.

\end{document}